\documentclass[12pt,a4paper]{conference}

\usepackage{fancyhdr}
\usepackage{graphicx,amsmath,amssymb,cite}
\usepackage{multind}
\makeindex{author} \makeindex{subject}

\newcommand{\beq}{\begin{equation}}
\newcommand{\eeq}[1]{\label{#1}\end{equation}}
\newcommand{\eeqn}{\end{equation}}

\newcommand{\beqa}{\begin{eqnarray}}
\newcommand{\eeqa}[1]{\label{#1}\end{eqnarray}}
\newcommand{\eeqan}{\end{eqnarray}}

\let\bar=\overbar

\newcommand{\Dslash}{\not{\hbox{\kern-4pt $D$}}}
\newcommand{\dslash}{\not{\hbox{\kern-2pt $\del$}}}

\newcommand{\msb}{{\bar{\ssstyle M \kern -1pt S}}}

\begin{document}

\Chapter{Hadron Physics with Anti-protons:\\The $\bar{\rm P}$ANDA Experiment at FAIR}
           {Hadron Physics with Anti-protons}{J.G.~Messchendorp}
\vspace{-6 cm}\includegraphics[width=6 cm]{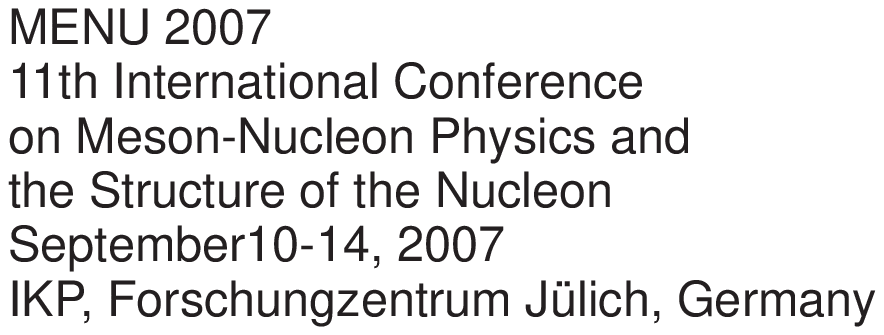}
\vspace{4 cm}

\addcontentsline{toc}{chapter}{{\it J.G.~Messchendorp for the $\bar{\rm P}$ANDA collaboration}} \label{authorStart}

\begin{raggedright}

{\it J.G. Messchendorp for the $\bar{\it P}$ANDA collaboration}\index{author}{Messchendorp, J.G.}\\
Kernfysisch Versneller Instituut\\
University of Groningen\\
Groningen, NL-9747 AA\\ 
The Netherlands \\
mail: messchendorp@kvi.nl
\bigskip\bigskip

\end{raggedright}

\begin{center}
\textbf{Abstract}
\end{center}

The theory of Quantum Chromo Dynamics (QCD) 
reproduces the strong interaction at distances much shorter 
than the size of the nucleon. At larger distance 
scales, the generation of hadron masses and confinement
cannot yet be derived from first principles on basis of QCD.
The ${\bar{\rm P}}$ANDA
experiment at FAIR will address the origin of these
phenomena in controlled environments. 
Beams of antiprotons together with a multi-purpose and compact 
detection system will provide unique tools to perform studies of 
the strong interaction. This will be achieved via precision 
spectroscopy of charmonium and open-charm states, an extensive 
search for exotic objects such as glueballs and hybrids, 
in-medium and hypernuclei spectroscopy, and more.
An overview is given of the physics program 
of the ${\bar{\rm P}}$ANDA collaboration.

\section{Introduction}

The fundamental building blocks of QCD are the quarks which 
interact with each other by exchanging gluons. QCD is well 
understood at short-distance scales, much shorter than the 
size of a nucleon ($<$~10$^{-15}$~m). In this regime, the basic 
quark-gluon interaction is sufficiently weak. In fact, many processes 
at high energies can quantitatively be described by perturbative QCD.
Perturbation theory fails when the distance among quarks becomes 
comparable to the size of the nucleon. Under these conditions, in the 
regime of non-perturbative strong QCD, the force among the quarks 
becomes so strong that they cannot be further separated. As a 
consequence of the strong coupling, we observe the relatively heavy 
mass of hadrons, such as protons and neutrons, which is two orders of 
magnitude larger than the sum of the masses of the individual quarks. 
This quantitatively yet-unexplained behavior is related to the 
self-interaction of gluons leading to the formation of gluonic 
flux tubes connecting the quarks. As a consequence, quarks have 
never been observed as free particles and are confined within hadrons, 
i.e. the baryons containing three valence quarks or mesons containing 
a quark-antiquark pair. 

\begin{figure}[hb]
\begin{center}
\includegraphics[width=10cm]{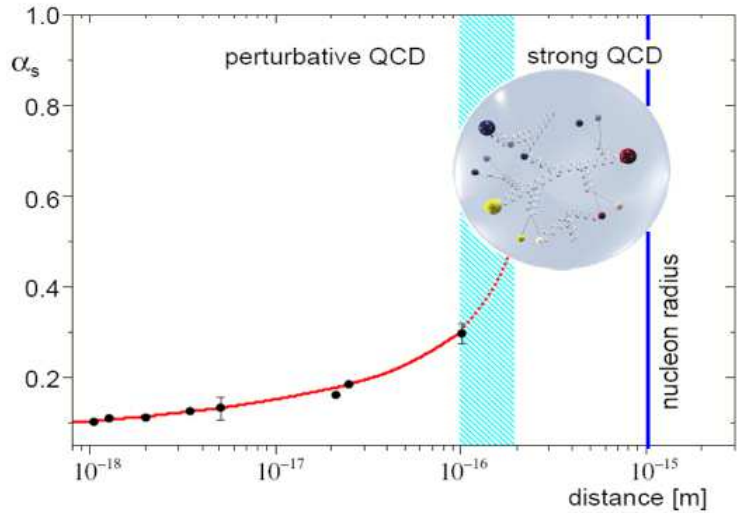}
\caption{\it The strong coupling constant, $\alpha_S$, as a function of the
distance scale. Towards larger distances, QCD becomes non-perturbative,
and gives rise to spectacular phenomena such as the generation of hadron 
masses.} \label{fig:alpha-panda}
\end{center}
\end{figure}

The physics program of the ${\bar{\rm P}}$ANDA (anti-Proton ANnihilation at DArmstadt)
collaboration~\cite{PANDA} will address various questions related to the strong 
interactions by employing a multi-purpose detector system at the 
High Energy Storage Ring for anti-protons (HESR) of the upcoming Facility 
for Anti-proton and Ion Research (FAIR)~\cite{FAIR}. The ${\bar{\rm P}}$ANDA collaboration aims
to connect the perturbative and the non-perturbative QCD regions, 
thereby providing insight in the mechanisms of mass generation and 
confinement. For this purpose, a large part of the program will be 
devoted to
\begin{itemize}
 \item charmonium spectroscopy;
 \item gluonic excitations, e.g. hybrids and glueballs;
 \item open and hidden charm in nuclei.
\end{itemize}
In addition, various other physics topics will be studied with ${\bar{\rm P}}$ANDA 
such as
\begin{itemize}
 \item the hyperon-nucleon and hyperon-hyperon interactions via 
$\gamma$-ray spectroscopy of hypernuclei;
 \item CP violation studies exploiting rare decays in the D and/or
$\Lambda$ sectors;
 \item studies of the structure of the proton by measuring Generalized Parton 
Distributions (Drell-Yan and Virtual-Compton Scattering), "spin" 
structure functions using polarized anti-protons, and electro-magnetic 
form factors in the time-like region.
\end{itemize}

\section{${\bar{\rm P}}$ANDA physics topics}

\begin{figure}[ht]
\begin{center}
\includegraphics[width=14cm]{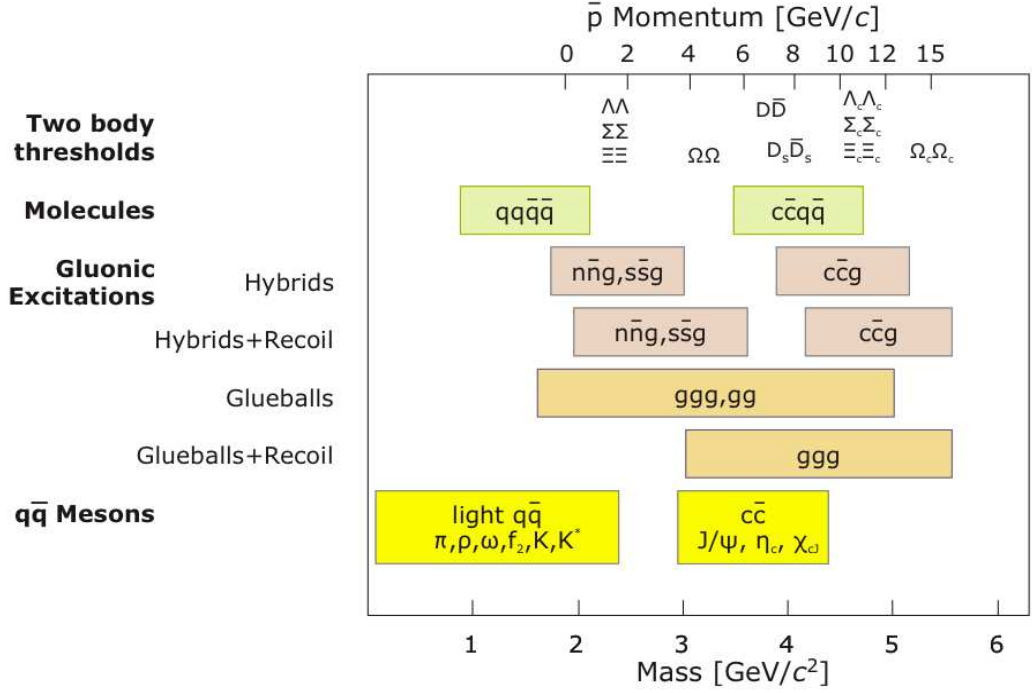}
\caption{\it An overview of the various quark and gluon configurations of 
hadrons and their corresponding mass range. The ${\bar{P}}$ANDA experiment will exploit 
masses up to 5.5~GeV/c$^2$ in antiproton-proton collisions, 
thereby having access to glueballs, charmed hybrids, and 
charmed-rich four-quark states.} \label{fig:meson-spectrum}
\end{center}
\end{figure}

The key ingredient for the ${\bar{\rm P}}$ANDA physics program is a high-intensity and 
a high-resolution beam of antiprotons in the momentum range of 1.5 to 15~GeV/c.
Such a beam gives access to a center-of-mass energy range from 2.2 to 5.5~GeV/c$^2$ 
in $\bar p$$p$ annihilations. In this range, a rich spectrum of hadrons with
various quark configurations can be studied as is illustrated in 
Fig.~\ref{fig:meson-spectrum}. In particular, hadronic states which contain 
charmed quarks and gluon-rich matter become experimentally accessible. 

\subsection{Charmonium Spectroscopy}

The level scheme of lower-lying bound $\bar c$$c$ states, charmonium, 
is very similar to that of positronium. These charmonium states
can be described fairly well in terms of heavy-quark potential models.
Precision measurements of the mass and width of the charmonium spectrum
give, therefore, access to the confinement potential in QCD.  
Extensive measurements of the masses and widths of the 1$^-$ $\Psi$ states 
have been performed at $e^+$$e^-$ machines where they can be formed directly
via a virtual-photon exchange. Other states, which do not carry the
same quantum number as the photon, cannot be populated directly, but only
via indirect production mechanisms. This is in contrast to the $\bar p$$p$ 
reaction, which can form directly excited charmonium states of 
all quantum numbers. As a result, the resolution in the mass and 
width of charmonium states is determined by the precision of the phase-space 
cooled beam momentum distribution and not by the (significantly poorer) 
detector resolution. 

The combination of the much better mass resolution 
with the ability to detect hadronic final states which have up to two 
orders of magnitude higher branching fractions than - for instance - 
the $\gamma\gamma$ decay channel will permit high-precision investigations
of charmonium states. The need for such a tool becomes evident by 
reviewing the many open questions in the charmonium sector.
For instance, our knowledge of the ground state, $\eta_c$, is 
surprisingly poor. The existing data~\cite{Par80,Him80,Bal86,Bis86,Bai00} 
do not present a consistent picture, 
and only a small fraction of the total decay width has been measured 
via specific decay channels. Furthermore, radial excitations, such as 
the $\eta_c'$, which was only recently discovered~\cite{Cho02}, are 
not simple recursions of the ground state, 
as was observed in the hadronic decays of the $\Psi$ states.
Another open question is the spin-dependence of the $q\bar q$ potential.
For this, a precise measurement of the mass and decay channels of the 
singlet-P resonance, $h_c$, is of extreme importance.
The available data for the $h_c$ are of 
poor precision~\cite{Ros05,Rub05}. 
Due to the narrow width, $\Gamma$$<1$~MeV, of this state, only 
$\bar p p$ formation experiments will be able to measure the width 
and perform systematic investigations of the decay modes. Finally, our
understanding of the states above the $D$$\bar D$ threshold is very poor 
and needs to be explored in more detail. Recent experimental
evidences (see review~\cite{Bar06}) hint at a whole series of surprisingly narrow states with masses
and properties which, so-far, cannot be interpreted consistently by theory.

Besides the spectroscopy of charmonium states, ${\bar{\rm P}}$ANDA will also provide
the capability to perform open-charm spectroscopy as the analog of
the hydrogen atom in QED (heavy-light system). Striking discrepancies
of recently discovered $D_{sJ}$ states by BaBar~\cite{Aub03} and CLEO~\cite{Bes03} 
with model calculations have been observed. Precision measurements of the
masses and widths of these states using antiprotons and by performing 
near-threshold scans are needed to shed light on these open problems.

\subsection{Hybrids, glueballs, and other exotics}

The self-coupling of gluons in strong QCD has an important consequence, 
namely that QCD predicts hadronic systems consisting of only gluons, 
glueballs, or bound systems of quark-antiquark pairs with a strong 
gluon component, hybrids. These systems cannot be categorized as 
"ordinary" hadrons containing valence $q\bar q$ or $qqq$. 
The additional degrees of freedom 
carried by gluons allow glueballs and hybrids to have spin-exotic 
quantum numbers, $J^{PC}$, that are forbidden for normal mesons 
and other fermion-antifermion systems. States with exotic quantum numbers
provide the best opportunity to distinguish between gluonic hadrons 
and $q\bar q$ states. Exotic states with conventional quantum 
numbers can be identified by measuring an overpopulation of the 
meson spectrum and by comparing properties, like masses, quantum numbers,
and decay channels, with - for instance - predictions from 
Lattice Quantum Chromodynamics (LQCD) calculations.

The first hints for gluonic hadrons came from antiproton
annihilation experiments. Two particles, first seen in $\pi N$ scattering
with exotics $J^{PC}$=1$^{-+}$ quantum numbers, $\pi_1$(1400)~\cite{Abe98} and 
$\pi_1$(1600)~\cite{Rei01} are clearly seen in $\bar p p$ at rest and are considered
as hybrid candidates. In the search for glueballs, a narrow state 
at 1500~MeV/c$^2$, discovered in antiproton
annihilations by the Crystal Barrel collaboration~\cite{Ams95,Ams95b,Abe96,Abe01} is considered the best candidate for
the glueball ground state ($J^{PC}$=0$^{++}$). However, the mixing
with nearby conventional scalar $q\bar q$ states makes a 
unique interpretation difficult. 

The most promising energy range to discover unambiguously 
hybrid states and glueballs is in the region of 3-5 GeV/c$^2$, 
in which narrow states are expected to be superimposed on a 
structureless continuum. In this region, LQCD predicts an exotic 1$^{-+}$ 
$\bar cc$-hybrid state with a mass of 4.2-4.5~GeV/c$^2$ and a glueball
state around 4.5~GeV/c$^2$ with an exotic quantum number 
of $J^{PC}$=0$^{+-}$~\cite{Mor99,Bal04}. The $\bar pp$ production cross section of 
these exotic states are similar to conventional states and in the
order of 100~pb. All other states with ordinary quantum numbers are
expected to have cross sections of about 1~$\mu$b.

\subsection{Hadrons in the nuclear medium}

One of the challenges in nuclear physics is to study
the properties of hadrons and the modification of these 
properties when the hadron is embedded in a nuclear 
many-body system. Only recently it became experimentally evident that 
the properties of mesons, such as masses of $\pi$, K , and $\omega$ mesons, 
change in a dense environment~\cite{Suz04,Bar97,Lau99,Trn05}.
The ${\bar{\rm P}}$ANDA experiment provides a unique possibility to extend 
these studies towards the heavy-quark sector by exploiting the $\bar p$$A$ reaction.
For instance, an in-medium modification of the mass of the $D$ meson would imply a modification of the
energy threshold for the production of $D$ mesons, compared to a free mass.
In addition, a lowering of the $D$-meson mass could cause
charmonium states which lie just below the $D\bar D$ threshold for the $\bar p p$ 
channel to reside above the threshold for the $\bar p A$ reaction. In such a case,
the width of the charmonium state will drastically increase, which can experimentally
be verified. Although this is intuitively a simple picture, in practice the situation
is more complicated since the mass of various charmonium states might also change
inside the nuclear medium. 

Besides the indirect in-medium studies as described above, ${\bar{\rm P}}$ANDA will 
be capable to directly measure the in-medium spectral shape of charmonium states. 
This can be achieved by measuring the invariant mass of the di-lepton decay products. 
For the $\Psi(3770)$, for instance, models predict mass shifts of the order 
of --100~MeV~\cite{Lee04}, which are experimentally feasible to observe.

\subsection{Hypernuclei}

Nuclei in which one or more of the constituent nucleons are replaced by hyperons, 
hypernuclei, are promising laboratories to study the hyperon-nucleon and 
hyperon-hyperon interactions. Single and double $\Lambda$-hypernuclei were 
discovered 50~\cite{Dan53} and 40~\cite{Dan63} years ago, respectively, of which only 6 double $\Lambda$-hypernuclei 
are presently known. With a dedicted setup of ${\bar{\rm P}}$ANDA and its antiproton beam, a copious 
production of double $\Lambda$-hypernuclei is expected to be observed, providing 
a precision investigation of the $\Lambda$-$\Lambda$ interaction. For this purpose,
antiprotons at a moderate momentum of 3~GeV/c will interact with a primary target
to produce large numbers of $\Xi\bar\Xi$ pairs. The $\bar\Xi$ decay provides a unique
signature for the production. The corresponding $\Xi$ particle will be stopped and
captured in a secondary target. In case the $\Xi$ is absorbed inside the nucleus, it will
yield a $\Lambda$ pair with very small (relative) energy. A (double) hypernucleus can be formed
whose decay is observed with spectroscopic precision using Germanium detectors. 

\subsection{Other Topics}

So far, this paper has concentrated on only a few of the topics which will be addressed
by the ${\bar{\rm P}}$ANDA collaboration. There exists, however, a large variety of other physics
topics which can ideally be studied with the ${\bar{\rm P}}$ANDA setup at the antiproton facility at FAIR. 

In one of these ``side'' activities, symmetry violation experiments are being proposed which
will open a window onto physics beyond the Standard Model of particle physics. This includes
experimental studies of lepton flavor number violation using rare decay of $D$ mesons and, 
in addition, CP violation studies by asymmetry measurements of ($D\bar D$) pairs 
near their production threshold in $\bar p p\rightarrow \Psi(3770)\rightarrow D\bar D$
and $\bar p p\rightarrow \Psi(4040)\rightarrow D\bar D$ reactions.

There is growing interest within the ${\bar{\rm P}}$ANDA collaboration to make use of electro-magnetic
probes, photons and leptons, in antiproton-proton annihilation. 
These probes will be used to study the structure of the proton by measuring 
Generalized Parton Distributions (GPDs), 
to determine quark distribution functions via Drell-Yan processes, 
and to obtain time-like electro-magnetic form factors by exploiting 
the $\bar pp\rightarrow e^+e^-$ reaction
with an intermediate massive virtual photon. For example, estimates~\cite{Fre03} of the 
count rates predict a few thousand $\gamma\gamma$
events per month for a luminosity of $2\cdot 10^{32}$~cm$^{-2}$s$^{-1}$ 
at an energy of $\sqrt{s}=3.2$~GeV/c$^2$ for the
crossed-channel Compton scattering process, which will be used to obtain GPDs.
This indicates that such studies are feasible using beams of antiprotons together
with a nearly-4$\pi$ electro-magnetic calorimeter, as foreseen with ${\bar{\rm P}}$ANDA.

\begin{figure}[ht]
\begin{center}
\includegraphics[width=12cm]{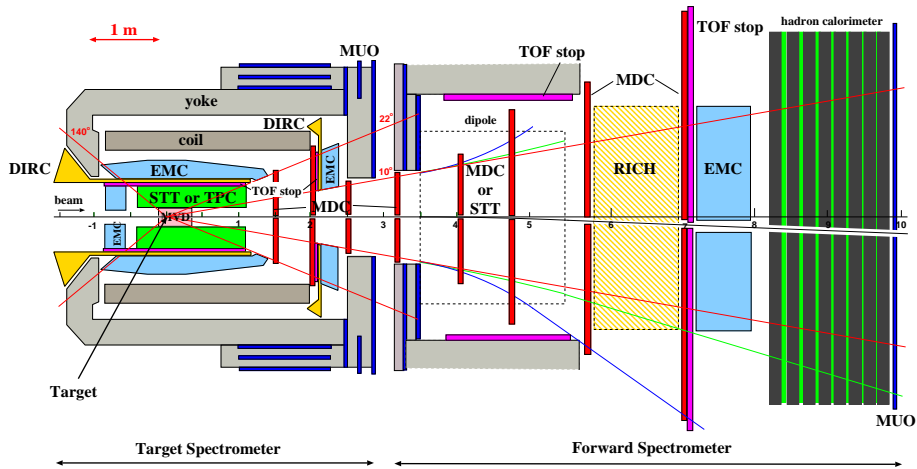}
\caption{\it A schematic side view of the ${\bar{P}}$ANDA detector. The different components are 
abbreviated as DIRC (Detection of internally reflected Cherenkov photons),
EMC (Electromagnetic calorimeter), STT (Straw-tube tracker), 
TPC (Time-projection chambers), MVD (Micro Vertex Detector), 
MDC (Mini drift chambers), MUO (Muon Detectors), 
RICH (Ring-imaging Cherenkov detectors), TOF (Time-of-flight detectors).
Not shown are the recent plans to include Gas Electron Multipliers (GEMs).} \label{fig:panda-setup}
\end{center}
\end{figure}

\section{The experimental facility}

The ${\bar{\rm P}}$ANDA detector will be installed at the High Energy Storage Ring, HESR,
at the future Facility for Antiproton and Ion Research, FAIR.  
FAIR provides a storage ring for beams of phase-space cooled antiprotons 
with unprecedented quality and intensity~\cite{HESR}. 
Antiprotons will be transferred to the HESR where internal-target experiments in 
the beam momentum range of 1.5 -- 15 GeV/c can be performed. Electron and stochastic 
phase space cooling will be available to allow for experiments with either 
high momentum resolution of about $\sim 10^{-5}$ at reduced luminosity or with 
high luminosity up to $2\times$10$^{32}$~cm$^{-1}$s$^{-1}$ with an enlarged momentum 
spread of $\sim 10^{-4}$.

The ${\bar{\rm P}}$ANDA detector is designed as a large acceptance multi-purpose setup. The 
experiment will use internal targets. It is conceived to use either pellets 
of frozen H$_2$ or cluster jet targets for the $\bar pp$ reactions, 
and wire targets for the $\bar pA$ reactions. 

In order to address the different physics topics, 
the detector needs to cope with a variety of final states and a large range of 
particle momenta and emission angles. 
At present, the detector is being designed to handle high rates of 10$^7$~annihilations/s , 
with good particle identification and momentum resolution for 
$\gamma,~ e,~ \mu,~ \pi,~ K$, and $p$ with the ability to measure 
$D,~ K^0_S$, and $\Lambda$ which decay at displaced vertices.  
Furthermore, the detector will have an almost 4$\pi$ detection coverage both for 
charged particles and photons. This is an essential requirement for an unambiguous partial 
wave analysis~\cite{Tim94} of resonance states. Various design studies are ongoing~\cite{Foe07}, 
partly making use of a dedicated computing framework for simulations and data 
analysis~\cite{Wro07}. A schematic overview of the detector is given in 
Fig.~\ref{fig:panda-setup}.

\section{Summary}

The ${\bar{\rm P}}$ANDA experiment at FAIR will address a wide range of
topics in the field of QCD, of which only a small
part could be presented in this paper. The physics program
will be conducted by using beams of antiprotons together with 
a multi-purpose detection system, which enables
experiments with high luminosities and precision resolution.
This combination provides unique possibilities to study 
hadron matter via precision spectroscopy of the 
charmonium system and the discovery of new hadronic matter, 
such as charmed hybrids or glueballs, as well as by measuring the properties
of hadronic particles in dense environments. New insights in the structure 
of the proton will be obtained by exploiting electromagnetic probes. 
Furthermore, the next generation of hypernuclei spectroscopy will be conducted 
by the ${\bar{\rm P}}$ANDA collaboration and at the J-PARC facility 
in Japan~\cite{JPARC}. 
To summarize, ${\bar{\rm P}}$ANDA has the ambition to 
provide valuable and new insights in the field of hadron physics which would bridge 
our present knowledge obtained in the field of perturbative QCD 
with that of non-perturbative QCD and nuclear structure.

\section*{Acknowledgments}

The author thanks N.~Kalantar-Nayestanaki, H.~L\"ohner, O.~Scholten, 
and R.G.E.~Timmermans for their comments and input to this paper. 
The ${\bar{\rm P}}$ANDA collaboration acknowledges 
the support of the European Community under the FP6 Design Study program. 
The present work has been performed with financial 
support from the University of Groningen and the Gesellschaft f\"ur 
Schwerionenforschung mbH (GSI), Darmstadt.

\end{document}